\documentclass[useAMS,usenatbib]{mn2e}  
\usepackage{amssymb,amsmath}
\usepackage{amsfonts}
\usepackage{graphicx}
\usepackage{booktabs}
\usepackage{tabularx}
\usepackage{color}

\numberwithin{equation}{section}

\def\d{{\rm d}}
\def\e{{\rm e}}
\newcommand{\p}{\partial}

\newcommand{\vw}{{\mbox{\boldmath $w$}}}

\newcommand{\case}{\textstyle\frac}

\newcommand{\eps}{\epsilon}

\newcommand{\vr}{{\mathbf r}}
\newcommand{\vx}{{\mathbf x}}
\newcommand{\vJ}{{\mathbf J}}
\newcommand{\vA}{{\mathbf A}}
\newcommand{\suml}{{\sum\limits}}
\newcommand{\il}{{\int\limits}}
\newcommand{\F}{{\cal F}}

\renewcommand{\Re}{\textrm{Re}\,}
\renewcommand{\Im}{\textrm{Im}\,}

\begin{document}

\title[Bisymmetric normal modes]{Bisymmetric normal modes in soft-centred and realistic galactic discs}
\author[E. V. Polyachenko \& A. Just]{E.~V.~Polyachenko, $^1 $\thanks{E-mail: epolyach@inasan.ru}\ \ \ A.Just, $^2 $\thanks{E-mail: just@ari.uni-heidelberg.de}\\
     $^1 ${\it\small Institute of Astronomy, Russian Academy of Sciences,} 
          {\it\small 48 Pyatnitskya St., Moscow 119017, Russia}\\
     $^2 ${\it\small Astronomisches Rechen-Institut, Zentrum f\"ur Astronomie der Universit\"at Heidelberg, } 
          {\it\small M\"onchhofstr. 12-14, 69120 Heidelberg, Germany}}



\maketitle

\begin{abstract}
We test methods for the determination of unstable modes in stellar discs: a point collocation scheme in the action sub-space, a scheme based on expansion of the density and potential on the biorthonormal basis, and a finite element method. Using models of galaxies with low and high mass concentration to the centre, the existence of two different kinds of spectra of unstable modes is demonstrated. Characteristic features of methods and obtained spectra are discussed.

Despite ignoring any constraint on the continuity or differentiability of the perturbed distribution function (DF), the collocation scheme is reliable for obtaining spectra and patterns in both kinds of models. The method based on the expansion of the perturbed potential and surface density over a biorthogonal basis was not applicable to a model with high mass concentration. The finite element method successfully used in various fields of science and engineering is currently sensitive to the presence of resonant orbits due to the choice of interpolation functions for the DF.

\end{abstract}

\begin{keywords}
galaxies: formation, galaxies: kinematics and dynamics.
\end{keywords}

\section{Introduction}

Modern theory of spiral structure has begun 50 years ago with works by \citet{LS64,LS66,K65}, but it is still not completed. One possible approach here is to find unstable eigenoscillations, or \textit{unstable modes}, of the stellar disc arising due to self-gravity. Unstable modes are characterized by their shape and complex frequency $\omega$, in which the real part determines the pattern speed of the spiral, $\Omega_\textrm{p} \equiv \Re \omega/m$, where $m$ is the number of arms, while the imaginary part gives the exponential growth rate of the amplitude, $\gamma \equiv \Im\omega $ (i.e. the growth rate of instability). If an unstable mode exists, a spiral pattern is formed spontaneously due to the growth of small fluctuations in the initially axisymmetric equilibrium distribution of the stars and the gravitational potential. Generally, there can be several unstable modes, and then the form of the spiral pattern is determined by the most unstable mode or a superposition of several modes in the case of close growth rates. The knowledge of the spectrum of unstable modes, i.e. their location in the ($\Omega_\textrm{p}, \gamma$)-plane, is obviously important for understanding the formation of the stellar disc and finding the shape of the spiral pattern.

This approach may be invaluable not only for the problem of spiral structure formation, but also for  the reconstruction of spiral patterns, when direct observations cannot be performed. Such a situation occurs in our own Galaxy.

In an alternative approach, the formation of spiral patterns is explained by the interplay of continuously arising, evolving, and fading transient spirals \citep{T81}. Unlike modes, which are waves with constant shape and frequency ($\Omega_\textrm{p}, \gamma$), they have a fast evolving form and frequency. The superposition of transient spirals can also lead to a quasi-stationary global picture.

Both approaches are not necessarily mutually exclusive. Depending on the shape of the rotation curve $v_\textrm{c}(r)$ and the phase space distribution function (DF), one or the other alternative can be realized. Moreover, they can coexist in one model \citep[e.g.][]{SC14}. However, if the disc has a large-scale unstable mode, then it should dominate and determine the shape of the global spiral pattern.

This paper considers available methods for finding disc unstable modes. For the first time, this problem was solved by \citet{K71,K77}. In linear perturbation theory, i.e. when the amplitude of oscillations is small compared to the corresponding unperturbed quantity, form and (complex) frequency of the mode are fixed. These characteristics can be formally obtained by the solution of a non-linear matrix equation. In practice, however, this Kalnajs matrix equation is effective only when a suitable set of potential--density basis functions is available. Moreover, due to the non-linearity of the equation a priori knowledge of the approximate values of the desired frequency is needed.

The linear matrix equation of the form
\begin{equation}
\vA\vx =\omega\vx\ ,
\label{lap}
\end{equation}
for unstable modes without using potential--density basis functions was suggested by \citet{EP05}. The advantage of his method is to obtain the entire spectrum immediately without a priori information on the localization of modes. However, the weak side of Polyachenko's method is the need to work with high-order matrices $\vA $. The question of the accuracy of the spectra and the individual modes is addressed in detail in Section 3.

Further efforts were directed to reduce the order of the matrix $\vA$. Using the Petrov-Galerkin projection for the collisionless Boltzmann equation, \citet{J07} proposed a method of the form (\ref{lap}), based on the introduction of a small number of `interpolating' functions in action sub-space and the expansion of the potential and surface density in a biorthogonal set of functions. The choice of the basis functions is limited to several sets \citep{CB72,K76,Q92,Q93}. In his next work \citet{J10} used the Bubnov--Galerkin weighting, where the expansion in a biorthogonal basis was replaced by a finite element method (FEM), which approximates the potential and the surface density in annular elements. This method is well established in structural engineering, as well as in various fields of physics.

Unstable solutions of kinetic equations can be of different nature. For example, the most unstable mode can occur due to a sharp jump of the DF at the disc edge, but be localized in the very centre. Hardly this mode can determine the shape of a global spiral pattern. Therefore, one needs to find all unstable modes, i.e.\ the entire spectrum.

Most of the available calculations of unstable modes are carried out for models with a smoothly rising rotation curve, so-called soft-centred models. Such models are characterized by low mass concentration to the centre, which account for the finite values of the angular velocity $\Omega$ and epicyclic frequency $\kappa$ in the centre, as well as the existence of a limiting value of the pattern speed when the spiral has no inner Lindblad resonance (ILR), $\Omega_\textrm{\,lim}\equiv\mathrm{max}\,(\Omega -\kappa/2)$. ILRs are crucial to the possibility of spiral and bar formation in the disc centre \citep{T69,M74} and thus for the shape of the spectrum of unstable modes. For soft-centred models, the modes form a wake pattern resembling one from a small boat bifurcating at $\Omega_\textrm{\,lim} $ \citep[e.g.][]{J07}.

In models with high mass concentration, e.g.\ with a flat circular velocity profile $v_\textrm{c}(r)$ or the profile based on observational data, the calculation of spectra is more complicated. The results are sensitive to uncertainties of model parameters derived from observations, and to parameters of the methods (computational meshes, etc.). Another problem relates to the interpretation of multiple `spurious' modes with small growth rates.

The aim of this work is to compare the methods by \citep{EP05,J07,J10} by calculating spectra of unstable bisymmetric ($m=2$) modes for several models of stellar discs with both smoothly rising and flat $v_\textrm{c}(r)$. In Section 2, we give a brief review of the linear matrix methods. Section 3 presents the models, spectra, and patterns, as well as discusses the applicability of the methods and the accuracy of the calculations. The  concluding section summarizes the results and outlines some perspectives.

\section{Matrix methods of linear perturbation theory}

In the linear approximation, we consider small perturbations of a system in equilibrium. Let the stars move in the plane of symmetry of the axially symmetric gravitational potential $V_0(r)$. The Hamiltonian $H_0$ depends only on the action variables $\vJ\equiv (J_r, L_z)$, which are integrals of motion. The equations of motion can be easily integrated, and the corresponding angle variables $\vw\equiv (w_1, w_2) $ are
\begin{equation}
w_1 =\Omega_1 (\vJ) t + w ^ 0_1\ ,\quad w_2 =\Omega_2 (\vJ) t + w ^ 0_2\ ,
\end{equation}
where $ (w^0_1, w^0_2) $ are some phases, and
\begin{equation}
\Omega_1 (\vJ)\equiv\frac{\p H_0 (\vJ)}{\p J_r}\ ,\quad\Omega_2 (\vJ)\equiv\frac{\p H_0 (\vJ)}{\p L_z}
\end{equation}
are the radial and azimuthal frequencies. In general both active (stellar disc) and passive (bulge, halo, etc.) components contribute to the  potential $V_0$. The circular velocity is determined by the total unperturbed potential
\begin{equation}
    v_\textrm{c} (r) =\sqrt{r\frac{\d V_0 (r)}{\d r}}\ .
\end{equation}

The unperturbed stellar disc is described by the equilibrium DF $F_0$ depending on the integrals of motion. A system near equilibrium is characterized by the perturbed DF and Hamiltonian
\begin{equation}
    F = F_0 (\vJ) + F_1 (\vJ,\vw, t)\ ,\quad H = H_0 + V_1 (\vr, t)\ ,
\end{equation}
where
\begin{equation}
    V_1 (\vr, t) = - G \int\d\vJ'\, \d\vw'\frac{F_1 (\vJ', \vw', t)}{|\vr -\vr'(\vJ',\vw')|}
   \label{poisson}
\end{equation}
($G$ is the gravitational constant) satisfying the collisionless Boltzmann equation:
\begin{equation}
\frac{\p F}{\p t} + [F, H] = 0\ .
\label{cbe}
\end{equation}
In the linear approximation, the problem is reduced to the solution of the linearized Boltzmann equation obtained by substituting in (\ref{cbe})
\begin{align}
 F_1 &=\e ^{imw_2 - i\omega t}\suml_{l = -\infty} ^{\infty}\F_l (\vJ)\e ^{ilw_1}\ , \notag\\
 V_1 &=\e ^{imw_2 - i\omega t}\suml_{l = -\infty} ^{\infty}\Psi_l (\vJ)\e ^{ilw_1}\ .
 \label{fourier}
\end{align}
Since the DF and the potential are related through the Poisson equation (\ref{poisson}), the ultimate problem is an eigenvalue problem in the form of an integro-differential equation.

\subsection{Linear matrix method (PME)}

\citet{EP05} proposed a matrix method in the form of a linear eigenvalue problem (\ref{lap}). In particular, for disc systems the equation is
\begin{multline}
\F_l (\vJ) [\omega - l\Omega_1 (\vJ) - m\Omega_2 (\vJ)] =\\
G F'_{0, l} (\vJ)\int\d\vJ '\suml_{l' = -\infty} ^{\infty}\Pi_{l, l '} (\vJ,\vJ ')\F_{l'} (\vJ ')\ ,
\label{bs}
\end{multline}
where $\{\F_l\}$ are eigenvectors specifying the shape of the spirals, $\F'_{0, l}(\vJ)$ denote a linear combination of partial derivatives with respect to the action variables
\begin{equation}
F'_{0, l} (\vJ) = l\frac{\p F_0 (\vJ)}{\p J_r} +
m\frac{\p F_0 (\vJ)}{\p L_z}\ .
\label{glbd}
\end{equation}
The kernels of the integral equations $\Pi_{l, l '} (\vJ,\vJ') $ can be represented as integrals over the radial angular variables
\begin{multline}
\Pi_{l, l '} (\vJ,\vJ') = 4\il_0 ^{\pi}\d w_1\il_0 ^{\pi}\d w'_1 h [r (w_1,\vJ), r (w'_1,\vJ ')] \times \\
\cos (l w_1 + m\theta)\cos (l' w'_1 + m\theta '),
\label{eq:pi}
\end{multline}
where
\begin{equation}
h [r (w_1,\vJ), r (w'_1,\vJ ')]\equiv\frac1{2\pi r_>}\il_0 ^{2\pi}
\frac{\cos (m\alpha)}{\sqrt{1 + z ^ 2 - 2z\cos\alpha}}\ ,
\end{equation}
$r(w_1,\vJ)$ is the radius of the stars on the unperturbed orbits, $z = r_</r_>$ is the ratio of the minimum and maximum radius with $r_< = \min [r(w_1,\vJ), r (w'_1,\vJ')]$, $r_> = \max [r(w_1,\vJ), r(w'_1,\vJ')]$;
\begin{equation}
\theta (w_1,\vJ) \equiv w_2 -\phi = w_2 - L_z\int^r_{r_{\min}} \frac{\d x}{x^2 v_r(x,\vJ)}\ ,
\end{equation}
 $\phi =\phi (w_1, w_2,\vJ) $ is a star's azimuth angle; $v_r$ is the radial velocity
\begin{equation}
v_r (r,\vJ) = [2 (E-\Phi (r)) - L_z ^ 2/r ^ 2] ^{1/2}\ .
\end{equation}

Equation (\ref{bs}) is valid for unstable modes with $\gamma\equiv\Im\omega> 0$ only. To find neutral and damped modes the \citet{L46} bypass rule must be applied. The integral in equation (\ref{bs}) is taken over the admissible region of action sub-space, which is a curvilinear triangle symmetric in $L_z$ and bounded by a line of circular orbits $J_r = 0$ and curves $J_\textrm{max}(|L_z|)$ corresponding to orbits with maximum energy of the stars.

From equation (\ref{bs}), one can obtain the angular momentum conservation law provided that $\Im \omega > 0$, $L_m = \sum_l L_m(l) = 0$, where $L_m(l)$ are Fourier components of angular momentum:
\begin{equation}
L_m(l)  = -\int\d\vJ F'_{0, l}(J) \frac{|\Psi_l (\vJ)|^2}{|\omega - l\Omega_1 (\vJ) - m\Omega_2 (\vJ)|^2}\ .
\label{am}
\end{equation}

In order to reduce (\ref{bs}) to the matrix equation (\ref{lap}), one must specify the grid in action sub-space ($\vJ$), and the cutoff parameters $(l_\textrm{min},  l_\textrm{max})$  for the Fourier expansions (\ref{fourier}). The most important terms in the expansions correspond to $l = -1,0,1$ \citep{EP04,EP05}. However, to get accurate results, one needs to take a larger range of summation on $l$ into account. Their total number $ N_l = l_\textrm{max}-l_\textrm{min} +1 $ depends on the model under investigation. The order of the matrix $\vA $ equals $ N_l N_\textrm{ph} $, where $ N_\textrm{ph} $ is the number of mesh points in action sub-space. For typical values of $ N_l\sim 10 $, $ N_\textrm{ph}\sim 1000$, which is common when using popular linear algebra software packages with modern PCs, approximately 1GB RAM is required to store the matrix $\vA$. The maximum order of the matrix in our calculations was $ 101\times12\times16 = 19392$.

\subsection{Linear method using basis functions (ECB)}

\citet{J07} proposed to take advantage of the linear eigenvalue problem of the previous method while reducing the order of the matrix $\vA$. To this end, instead of equation (\ref{fourier}) he introduced a decomposition
\begin{equation}
\F_l(\vJ) =\suml_{j=0}^{\infty} d_j^l\Phi^l_j(\vJ)\ ,\quad
 V_1 =\suml_{j=0}^{\infty} b_j^l\Psi^l_j(\vJ)\ ,
 \label{J07fourier}
\end{equation}
where $\Phi^l_j$, $\Psi^l_j $ are interpolating functions. In coordinate space, the perturbed surface density and the potential are represented as the real part of the expansion in biorthogonal basis functions
\begin{align}
  V_1(r,\phi,t) &=\e^{im\phi - i\omega t}\suml_{j=0}^{\infty} a_j\psi_j (r)\ ,\notag\\
\Sigma_1(r,\phi,t) &=\e^{im\phi - i\omega t}\suml_{j=0}^{\infty} a_j\sigma_j (r) \ ,
 \label{J07bobe}
\end{align}
which satisfies the orthogonality condition
\begin{equation}
 2\pi\il_0^\infty\psi_j(r)\sigma_{j'} (r) r\d r = D_j\delta_{j,j'}\ ,
 \label{J07bob}
\end{equation}
where $\delta_{j, j'}$ is Kronecker delta, and $D_j$ are normalization constants.

The interpolating functions for the potential can be conveniently chosen as
\begin{equation}
\Psi_j(\vJ) =\frac1\pi\il_0^\pi\psi_j(r)\cos[l w_1 + m(w_2 -\phi)] dw_1\ .
\end{equation}
The choice of interpolating functions for the DF is less obvious. \citet{J07} proposed to adopt it in the form
\begin{equation}
\Phi_j(\vJ) =\frac{F'_{0,l}}{l\Omega_1 + m\Omega_2}\Psi_j(\vJ)\ ,
\label{J07Phi}
\end{equation}
where $F'_{0,l}$ is given in equation (\ref{glbd}). Note, however, that this choice has a potential problem for systems with resonant orbits
\begin{equation}
 l\Omega_1(\vJ) + m\Omega_2(\vJ) = 0\ .
\end{equation}
This condition holds for any radial orbit and any orbit in the harmonic potential when $(l, m) = (-1,2)$.
Due to this resonance, equation 41 in \citet{J07} for the correction of the matrix equations in the presence of a sharp cutoff in the DF on radial orbits is undefined.

A choice of the biorthogonal basis is also a problem. As was noted in the introduction, it is limited to a few sets. Following \citet{J07}, we use the \citet{CB72} basis functions for this method (referred as ECB).

\subsection{Finite element method (FEM)}

In the subsequent paper, \citet{J10} abolished the basis functions (\ref{J07bobe}) in favour of the FEM, that implies expansion of $V_1$ and $\Sigma_1$ in polynomial functions inside ring elements. Let $r_n$, $n = 1,\dots,N$, be a set of nodes along the radius. As above, the perturbation of the surface density and the potential is determined by the real part of the functions $V_1(r,\phi, t) $ and $\Sigma_1(r,\phi, t)$, but now
\begin{align}
 V_1 (r,\phi, t) &=\e^{im\phi - i\omega t}\suml_{n = 1} ^{N} \suml_{j = 1} ^{N_d} H_n(r) G_n^j a^j_n\ ,\notag\\
\Sigma_1 (r,\phi, t) &=\e ^{im\phi - i\omega t}\suml_{n = 1} ^{N} \suml_{j = 1} ^{N_d} H_n(r) G_n^j b^j_n\ ,
 \label{J10e}
\end{align}
where the functions $H_n(r)$ equal unity inside the ring $r_n\le r\le r_{n +1}$ and zero otherwise. The interpolating polynomials $G_n^j$, $1\le j \le N_d$ must satisfy $G^j_n (\bar r_k) = \delta_{jk} $, where $\bar r_k$ is the radial position of the $k$th node in the normalized coordinates:
\begin{equation}
\Bar r = 2\frac{r-r_n}{\Delta r_n} - 1\ .
\end{equation}
The simplest case corresponds to $N_d = 2$:
\begin{equation}
G ^ 1_n =\frac12 (1 -\bar r)\ ,\quad G ^ 2_n =\frac12 (1 +\bar r)\ .
\end{equation}
  
The Fourier components of the perturbed potential and the DF are represented by
\begin{align}
   \Psi_l (\vJ) =\suml_{n = 1} ^{N}\suml_{j = 1} ^{N_d}\Psi_l ^ j (n,\vJ) a ^ j_n\ ,\notag\\
   \F_l (\vJ) =\suml_{n = 1} ^{N}\suml_{j = 1} ^{N_d} E_l ^ j (n,\vJ) z_l ^ j (n)\ 
\end{align}
with interpolating functions for the potential
\begin{equation}
   \Psi_l ^ j (n,\vJ) =\frac{1}{2\pi}\il_0 ^\pi H_n (r) G_n ^ j\cos [ilw_1 + im (w_2-\phi)]\d w_1\ .
\end{equation}
The choice of the interpolating functions for the perturbed DF $E_l ^ j (n,\vJ)$ is crucial for the performance of the method. \citet{J10} took them in the form
\begin{equation}
    E_l ^ j (n,\vJ) =\frac{F'_{0, l} (\vJ)}{l\Omega_1 + m\Omega_2}\Psi_l ^ j (n,\vJ)
\label{J10Phi}
\end{equation}
similar to equation (\ref{J07Phi}). Such a choice obviously contains the same trouble with resonance orbits as the ECB method.

 \citet{J10} offered two types of matrix equations: the full method suitable for arbitrary models (equation 58 in the cited article), and an approximate method for discs mostly populated by nearly circular orbits (see 63 ibid.). In the second method, one neglects the long-range interactions by means of trajectory deformations between different annular elements. In this paper, we denote these methods as FEMf and FEMc, respectively.

\section{Shapes of the spectra and spiral patterns}

The methods described in Section 2 are compared using three models of galaxies. Two of them, the exponential disc and the Mestel disc, are well known in the literature. In the exponential disc model, we demonstrate the convergence of all methods, and discuss characteristic features of the spectra and spiral patterns. For the Mestel disc, the well-known Toomre--Zang model is used \citep{Z76,T77}, for which the frequency of the single unstable mode is known with very high accuracy. The flat circular velocity profile $v_\textrm{c}(r)$ of the Mestel disc fits better to describe galactic rotation curves, but brings difficulties in the centre of the disc, where the frequencies $\Omega_{1,2}$ grow indefinitely and therefore the ILR occurs for any value of $\Omega_\textrm{p}$.

The third model is our first attempt to study the spectral properties of galactic disc models with realistic rotation curves. The mass model here consists of a disc and a halo component taken from the three-component model of the Galaxy by \citet{GJ13}. In this example, we illustrate the problem of resonant denominators of the interpolating functions (\ref{J07Phi}) and (\ref{J10Phi}) and we offer a way to resolve it.

In this paper, we restricted ourselves to the study of $m = 2$ modes, i.e. bar modes and two-armed spirals. Also, we will not consider questions related to the presence of radial orbits. In all our models stars rotate in one direction and the DF vanishes for purely radial orbits.

The orbits in action sub-space are conveniently described by ($R_c, e$), which replaces $\vJ$ after three successive changes of variables \citep{J10}:
\begin{equation}
  (J_r, L_z)\to (E, L_z)\to (R_\textrm{min}, R_\textrm{max})\to (R_c, e)\ ,
 \label{grid}
\end{equation}
where $R_\textrm{min}$ and $R_\textrm{max}$ are pericentric and apocentric radii
\begin{equation}
  R_\textrm{min} = R_c (1-e)\ ,\quad R_\textrm{max} = R_c (1 + e) \ ,\quad 0\le e\le 1\ .
 \label{grid1}
\end{equation}
The mesh for $R_c$ is determined by parameters $R_c ^\textrm{min}$, $R_c ^\textrm{max}$ and $N_J$:
\begin{align}
  R_c^j &= \exp (u_j) - R_\textrm{min}\ ,\notag\\
  u_j   &= \frac{u_\textrm{max} - u_\textrm{min}}{N_J-1} (j-1) + u_\textrm{min}\ ,\quad j = 1,\dots,N_J\ ,
 \label{grid2}
\end{align}
where $ u_\textrm{min} =\ln (2 R_c ^\textrm{min}) $, $ u_\textrm{max} =\ln (R_c ^\textrm{min} + R_c ^\textrm{max}) $. The grid for the eccentricity $e$ with number of nodes $N_e$ is uniform:
\begin{equation}
  e_j = (j-\case12)/N_e\ ,\quad j = 1,\dots,N_e\ .
 \label{grid3}
\end{equation}

The cutoff parameters $(l_\textrm{min}, l_\textrm{max})$ in the Fourier series (\ref{fourier}) for soft-centred models are $(-5,5)$. For the Toomre--Zang model the convergence on these parameters is investigated separately.

\subsection{The exponential disc}

In this model we assume a soft-centred logarithmic potential
\begin{equation}
  V_0 (r) = v_0 ^ 2\ln\sqrt{1 + r ^ 2/r_C ^ 2}\ ,
 \label{vc_cem}
\end{equation}
which describes a linearly increasing circular velocity for $ r\ll r_C $, and a nearly flat one reaching $v_0$ at $r \gg r_C$. Hereafter, we assume units in which $ G = v_0 = r_C = 1$, leading to $\Omega(r) \le 1$ for the angular velocity. The limiting frequency is $\Omega_\textrm{lim}\simeq 0.106 $. The surface density of the disc is
\begin{equation}
 \Sigma_D (r) =\Sigma_s\exp\Big [-\lambda\sqrt{1 + r ^ 2/r_C ^ 2}\Big]\ ,\quad\lambda = \frac{r_C}{r_D}\ ,
 \label{exp_sd}
\end{equation}
where $ r_D $ is the disc scalelength.

\begin{figure} 
  \centerline{\includegraphics [width = 85mm]{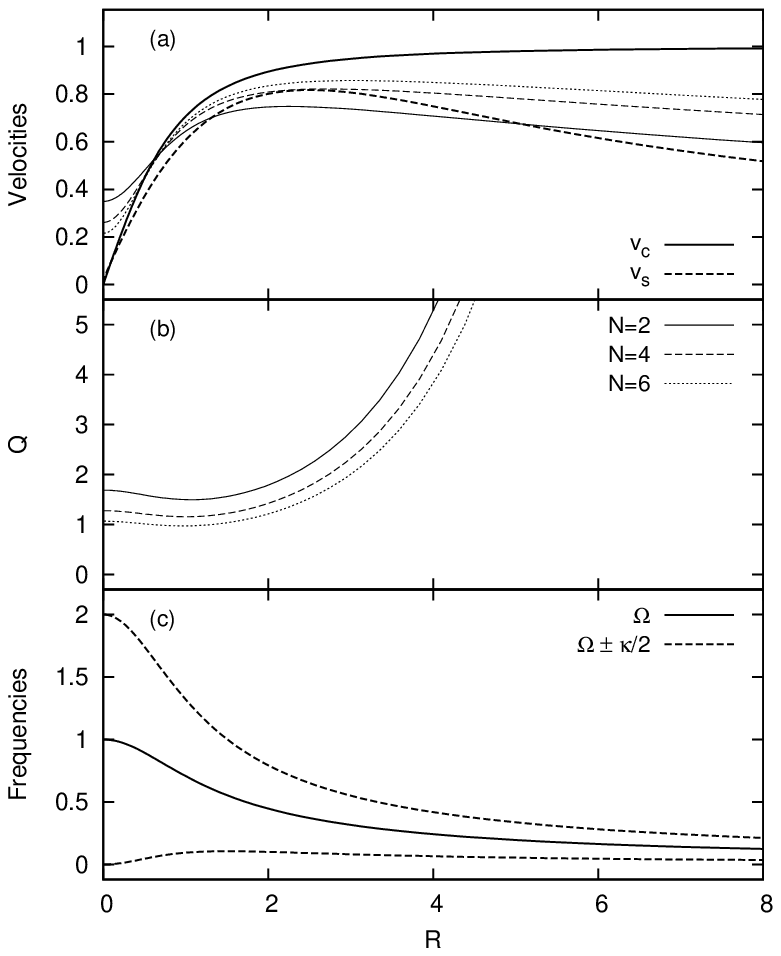}}
  \centerline{\includegraphics [width = 85mm]{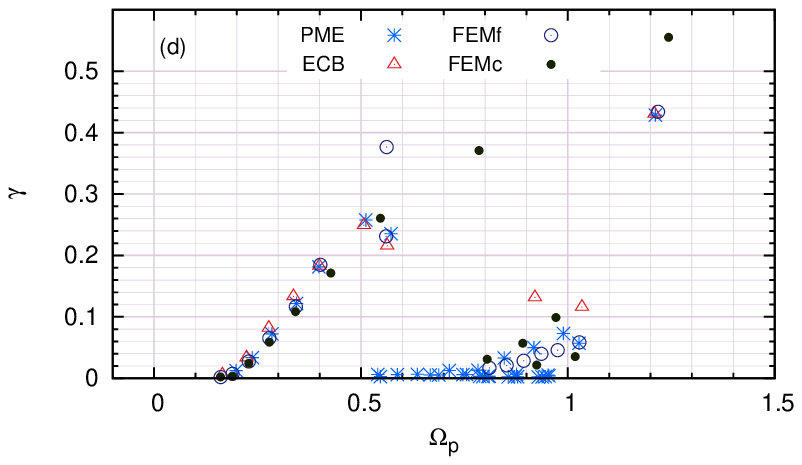}}
  \caption{The exponential disc with $\lambda = 1 $, $\Sigma_s = 0.42 $: (a) Circular velocity $v_\textrm{c}(r) $ for the potential (\ref{vc_cem}) and circular velocity $v_s(r)$ due to the disc component (\ref{exp_sd}) only; thin curves show the profiles of the mean rotation velocity depending on $N$; line types are the same as in (b). (b) Toomre's $Q$ for $ N = 2,4,6 $. (c) Angular velocity $\Omega(r)$ and curves $\Omega(r) \pm \kappa(r)/2 $, determining the position of Lindblad resonances. Maximum  $\Omega -\kappa/2 $ of the precession velocity is $\Omega_\textrm{lim} = 0.106 $. 
     (d) Spectra of unstable modes in the ($\Omega_\textrm{p},\gamma$)-plane for the exponential model $N = 6$ obtained by PME, ECB, FEMf, and FEMc.}
  \label{fig1}
\end{figure}

The DF of this model contains a free integer parameter $N$ \citep[e.g.][]{JH05}. Besides, the cutoff function $ H_\textrm{cut} = 1 -\exp (-L_z ^ 2/L_0 ^ 2) $ is applied in order to remove stars with small angular momenta. Thus, the model has four parameters: $(N,\lambda,\Sigma_s, L_0) $. We consider the model $ (N, 1,0.42,0.1) $, in which the potential is mainly determined by the disc component (`maximum' disc, see Fig.\,\ref{fig1}(a)). With increasing $N$, the disc becomes colder as it follows from Toomre's $Q$ profiles shown in Fig.\,\ref{fig1}(b):
\begin{equation}
  Q =\frac{\kappa\sigma_r}{3.36G\Sigma_D }\ ,
\end{equation}
where $\kappa(R_c) = \Omega_1(R_c,e)|_{e=0}$ is the epicyclic frequency, $\sigma_r (r)$ is the radial velocity dispersion.

Fig.\,\ref{fig1}(a) also shows the profiles of the mean rotation velocity $\langle v_\phi\rangle $, which approach circular velocity $v_\textrm{c}$ with increasing $N$. Fig.\,\ref{fig1}(c) shows the angular velocity $\Omega(r)$ and curves $\Omega(r)\pm\kappa (r)/2$, which provide the location of the main resonances for a given pattern speed $\Omega_\textrm{p}$. If $\Omega_\textrm{p}>\Omega_\textrm{lim} $, then the ILR is absent, thus bar and spiral modes reaching to the very centre must have $\Omega_\textrm{p}> 0.106$.

Hereafter we consider the $N=6$ case only. The outcome of the matrix methods is given in Fig.\,\ref{fig1}(d). Eigenvalues of the matrix $\vA$ include both purely real solutions and complex conjugate pairs. The unstable modes we are interested in are localized in the upper half-plane $\gamma\equiv\Im\omega> 0 $. Note that all methods give a number of modes with very small growth rates $\gamma$, which are indistinguishable from zero. Therefore, in the figure we have restricted ourselves to modes with $\gamma > 0.002$. The modes are grouped in two regions. The first group with $\Omega_\textrm{p}\sim 1$ contains modes localized in the centre. The length of the spiral patterns increases with the decrease of $\Omega_\textrm{p}$. For modes of $\Omega_\textrm{p} > 1 $ the corotation resonance (CR) is absent. The second group of modes bifurcates from $\Omega_\textrm{p} =\Omega_\textrm{lim}$ to the right up to $\Omega_\textrm{p}\sim 0.6$. Modes with $\Omega_\textrm{p} <\Omega_\textrm{lim} $ have ILRs. However, our calculations do not show the presence of such modes, as well as modes with $\Omega_\textrm{p}> 1.3$.

The frequency of the most unstable mode of the first group $ (\Omega_\textrm{p},\gamma)\approx (1.21, 0.43) $ is well defined by all methods, except the approximate FEMc. For other modes in the group, the coincidence is less impressive: close results for the next two strongest (with the highest growth rates) modes give PME and FEMf, while the two other methods give significantly different results.

\begin{figure*}
  \centerline{\includegraphics [width = 170mm]{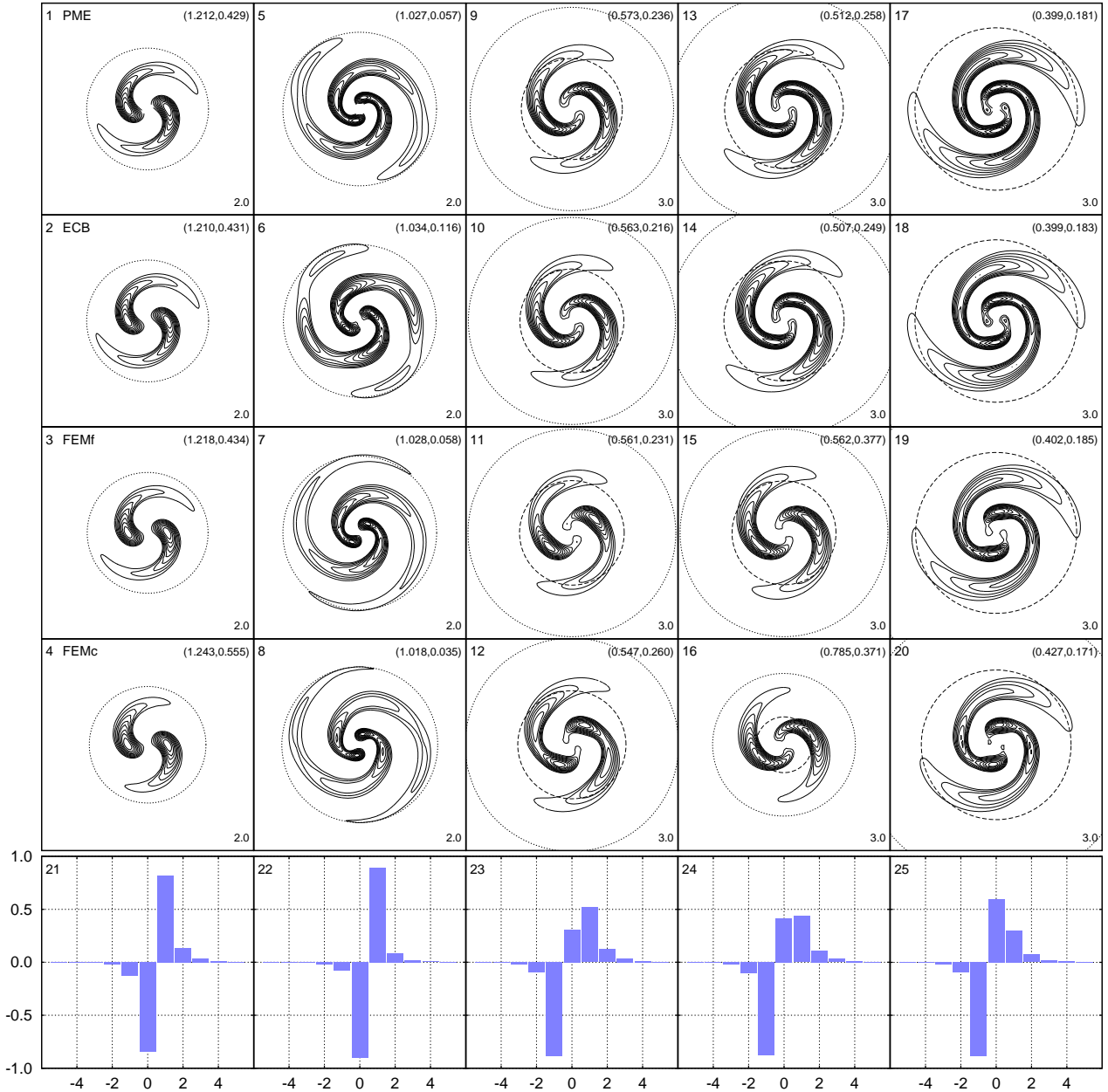}}
  \caption{Spiral patterns of the exponential model $N = 6$, obtained by PME, ECB, FEMf, FEMc (rows, respectively). Isolines show the excess surface density over the axially symmetric distribution by 10--90 per cent incrementing by 10 per cent. The pattern speeds and the growth rates are given in parentheses, $(\Omega_\textrm{p},\gamma)$. Dashed lines show CR, while dotted lines  show the Lindblad resonances. Frames (1--8) contain patterns for the first group of modes ($\Omega_\textrm{p}\gtrsim 0.8$), frames (9--20) -- for the second group of modes, $\Omega_\textrm{p}\lesssim 0.6$. The pattern frame sizes are shown in the lower right corner. The bar charts indicate the dependence $L_2(l)$ versus $l$ reflecting angular momentum exchange between different parts of the disc.}
  \label{fig3}
\end{figure*}

The spiral patterns obtained by the different methods can be compared by the radial extend, the pitch angle, and the number and location of the maxima. The patterns rotate counterclockwise so all spirals are trailing. The spiral patterns of the first group are shown in Fig.\,\ref{fig3}. The pattern speed and the growth rate $ (\Omega_\textrm{p},\gamma)$ are provided for each pattern. Patterns for the most unstable modes (frames 1--4) are almost identical: they have one maximum at $ r\approx 0.38 $, the same pitch angle, and almost equal length of the spirals. For the second modes (frames 5--8), similar spirals were obtained by PME, FEMf, and FEMc, while the ECB gives a spiral with changing pitch angle and larger number of maxima.

All methods equally well reproduce the frequencies of modes with small $\gamma$ localized near $\Omega_\textrm{lim} $. However, the most unstable modes of this group are determined inaccurately. PME and ECB show the best match: they repeat the entire shape of the spectrum, including a somewhat smaller growth rate for the fastest mode. The FEMf method also shows a nonmonotonic dependence of $\gamma $ with $\Omega_\textrm{p}$. However, the growth rate of the most unstable mode of this group greatly exceeds its counterparts. Unlike the other methods, FEMc modes monotonically bifurcate from the $x$-axis, and instead of a pair of FEMf modes (0.561, 0.231), (0.562, 0.377) one obtains the FEMc pair (0.547, 0.260) and (0.785, 0.371), i.e. the pattern speed of one of the modes is overestimated by 50 per cent.

Frames 9--20 in Fig.\,\ref{fig3} presents spiral patterns of modes of the second group. Patterns of the PME and ECB modes are indistinguishable. The number of peaks and the extent of spirals increase with decreasing $\Omega_\textrm{p}$. All methods give almost the same patterns for modes with minimum growth rates (frames 17--20). The difference is noticeable between the patterns (11), (12) and (15), (16).

The bar charts (frames 21--25) show loss and gain of angular momentum by different Fourier components in the direction of radial angle (\ref{am}). For low growth rates, physical meaning of $L_m(l)$ is loss or gain of angular momentum on the resonance $(l,m)$: $l=0$ corresponds to CR, $l=\pm 1$ -- to Lindblad resonances. The bars are normalized so that the sum of positive components is unity. The sum $\sum_l L_m(l)$ must vanish because the total angular momentum of the disc is conserved. PME and ECB schemes obey the conservation law perfectly (typical errors are $\sim 10^{-9}$ and $\sim 10^{-15}$, respectively), and the bars obtained are identical. On the contrary, calculation of FEM bars gives typical errors 10--50 per cent, and we reject them from the charts. Throughout the paper, the bars reflect angular momentum exchange for PME modes.

For modes of the first group $\Omega_\textrm{p}\gtrsim 0.8$ angular momentum is lost by the $l \leq 0$ Fourier components, primarily by $l=0$, and gained by the $l \geq 1$ components. Since  the modes lack CRs, the angular momentum transfer for the very centre of the disc to the outer Lindblad resonance (OLR). For modes of the second group $\Omega_\textrm{p}\lesssim 0.6$, $l=0$-components are positive, implying the exchange from the centre to CR and OLR.

A run of the PME method with mesh parameters $ (N_J, N_e) = (101, 8) $ on a standard PC takes less than 15 min: 3.5 min for the calculation of the nuclei elements (\ref{eq:pi}), and 11.3 min for the calculation of eigenvectors and eigenvalues of the matrix $\vA$. The matrix order in this case was $8888$. For the same parameters, ECB with a number of potential--density pairs $j_\textrm{max} = 15$ took only 24 s (the matrix order was $176$), while both FEM methods with $ (N, N_d) = (100, 2)$ took 2.5 min (the matrix order was $1111$).

\subsection{Toomre--Zang model}

Unlike most models used for the stability study, which have a smoothly rising circular velocity profile, this model is characterized by a flat profile, and thus the ILR is present for any value of the pattern speed. Following \citet{T77}, we adopt an unperturbed DF of energy $ E $ and angular momentum $L_z$
\begin{equation}
  F_0 (E, L_z) = C\Big (\frac{L_z}{r_0 v_\textrm{c}}\Big) ^ q{\rm e} ^{-E /\sigma ^ 2}
 \label{mestel_df}
\end{equation}
for $ L_z> 0 $ and zero for $ L_z <0 $, which determines a self-consistent Mestel disc
\begin{equation}
 \Sigma_D (r) =\Sigma_0\frac{r_0}{r}\ ,\quad V_0 (r) = v_0 ^ 2\ln\frac{r}{r_0}\ ,
 \label{mestel}
\end{equation}
with $ v_\textrm{c} = v_0 = \sqrt{2\pi G\Sigma_0 r_0} $ and radial velocity dispersion $\sigma = \textrm{const}$ provided by
\begin{equation}
 q\equiv\frac{v_0 ^ 2}{\sigma ^ 2} - 1\ ,\quad C =\frac{\Sigma_0 v_0 ^ 2}
{2 ^{q/2}\sqrt{\pi} (\frac12 q -\frac12)!\sigma ^{q +2}}\ .
\end{equation}
The cutoff function
\begin{equation}
 H_\textrm{cut} =\frac{1}{1 + (L_0/L) ^ n}\ ,\quad L_0 = v_0 r_0
\end{equation}
($ n $ -- integer) turns the DF smoothly to zero at $ L_z=0 $.

According to \citet{T77}, the unstable modes are observed only for sufficiently large $ n $, i.e. they are due to the sharp edge of the cutoff. In units $ v_0 = G = r_0 = 1$, for $ n = 4 $ and $ q = 6$  the exact value of the single unstable mode is $(\Omega_\textrm{t},\gamma_\textrm{t}) = (0.439426, 0.127181) $ (test frequency).

\begin{figure}
  \centerline{\includegraphics [width = 85mm]{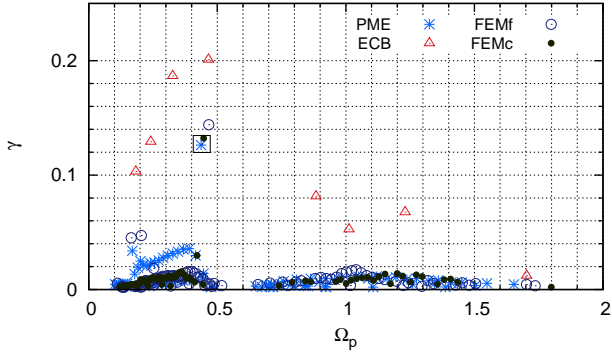}}
  \caption{Spectra of the Toomre--Zang model obtained by the various methods. The square indicates the position of the test frequency $ (\Omega_\textrm{t},\gamma_\textrm{t}) = (0.439426, 0.127181) $. }
  \label{fig5}
\end{figure}

Fig.\,\ref{fig5} shows the spectra obtained by all methods under test. We used $ (l_\textrm{min}, l_\textrm{max}) = (-5, 10)$ in all calculations. For PME $ (N_J, N_e) = (151,8)$, whereas in the others $ (N_J, N_e) = (201,64)$; other parameters were $ N = 200 $, $ N_d = 2 $, $ j_\textrm{max} = 15 $. As in Fig.\,\ref{fig1}(d), we do not show modes with growth rates $\gamma <0.002 $.

PME, FEMf, and FEMc spectra contain one mode with growth rate above 0.1; PME and FEMc give $\Omega_\textrm{p}$ and $\gamma$ that agree well with the test values, while FEMf overestimates these parameters. The ECB method provides many unstable modes in the region of interest, and thus is not applicable to this model.

The spectra include spurious modes grouped in two sets, growth rates of which are sensitive to a change of the mesh parameters. Their growth rates in our best calculations were at the level $\langle\gamma_s\rangle\lesssim 0.02 $. The level depends largely on the choice of the mesh and the number of nodes. For example, when using meshes from \citet{J10}, growth rates of some modes are comparable or even larger than $\gamma_\textrm{t}$. Lower values of $\langle\gamma_s\rangle $ can be achieved by increasing the number of nodes $ N_J $ (PME), or by increasing the number of rings $ N $ (FEM): $\langle\gamma_s\rangle $ is approximately inversely proportional to these parameters. Changing $ N_d $ has no effect on $\langle\gamma_s\rangle $.

\begin{figure*}
  \centerline{\includegraphics [width = 170mm]{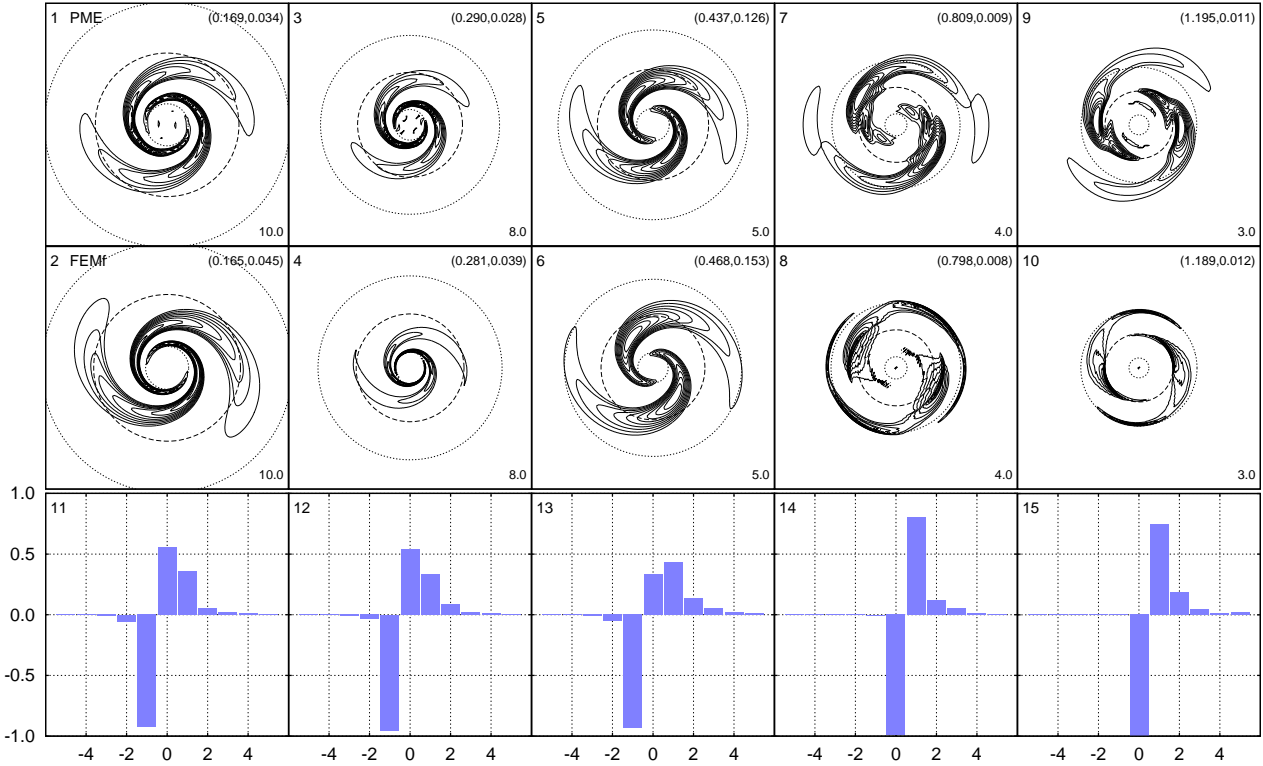}}
  \caption{Same as in Fig.\,\ref{fig3} for the Toomre--Zang model, obtained by PME and FEMf. Frames 1,2,5,6 -- the edge modes; 3,4 -- examples of the patterns for spurious modes of set 1 ($\Omega_\textrm{p} \lesssim 0.4$); 7--10 -- examples of the patterns for spurious modes of set 2 ($\Omega_\textrm{p} \gtrsim 0.65$). }
  \label{fig_tzp}
\end{figure*}

The first set extends from $\Omega_\textrm{p}\approx 0.17$, for which OLR is on the outer edge of the mesh, to $\Omega_\textrm{p}\approx 0.4$. Two modes-outliers with $\gamma \approx 0.04$ are obviously the edge modes due to the outer edge. The patterns of the edge modes (both outer-edge and inner-edge Toomre--Zang modes) are regular and occupy considerable part of the disc, see frames 1,2 and 5,6 in Fig.\,\ref{fig_tzp}. According to bar charts 11 and 13, angular momentum here is transferred from the region near ILR to basically CR and OLR.

The second set extends from $\Omega_\textrm{p}\approx 0.65$ to $1.7$. The right boundary is determined by position of the OLR near the inner edge of the disc $R_\textrm{OLR} \sim L_0/v_0$. The gap between the sets is formed when ILR moves outside the disc region, $R_\textrm{ILR} \lesssim L_0/v_0$. This is confirmed by examination of the bar charts of angular momentum exchange (frames 11--15).

Fig.\,\ref{fig_tzp} also shows some patterns of spurious modes. Frames 3,4 are examples from the first set; all patterns of this set are regular and extend from the ILR to the CR. The patterns of modes from the second set are less regular, as it follows from frames 7--10, and extend from the CR to the OLR.

Expanding the range of summation on $ l $ in the Fourier decomposition (\ref{fourier}) leads to an increase in $ (\Omega_\textrm{p},\gamma) $ of the PME mode, which tend to the values given by Toomre and Zang. Both, pattern speed and growth rate of FEMc mode, are also close to $ (\Omega_\textrm{t},\gamma_\textrm{t}) $, but slightly exceed it. At the same time  FEMf significantly overestimates the test values (see Table 1). Increasing the range of summation on $ l $ from $-5,\dots,10$ to $-10,\dots,10$ changes the frequency only in the fourth digits. Table 2 shows the frequency of the fundamental PME mode for different mesh parameters.

\begin{table}
\begin{center}
\begin{tabular}{l l l l}
\hline
 $ l $ & PME & FEMf & FEMc\\
\hline
  $-2,\dots,5$ & (0.4331, 0.1158) & (0.4651, 0.1405) & (0.4390, 0.1248)\\
  $-5,\dots,5$ & (0.4354, 0.1182) & (0.4669, 0.1440) & (0.4432, 0.1250)\\
  $-5,\dots,10$ & (0.4371, 0.1260) & (0.4683, 0.1531) & (0.4470, 0.1322)\\
  $-10,\dots,10$ & (0.4372, 0.1261) & (0.4685, 0.1534) & (0.4473, 0.1321)\\
\hline
\end{tabular}
\end{center}
\vspace{-2mm}
\caption{Convergence of frequencies to the test values $ (\Omega_\textrm{t},\gamma_\textrm{t}) = (0.439426, 0.127181) $ when changing the cutoff in the Fourier decomposition (\ref{fourier}): PME uses $ (N_J, N_e) = (101, 8)$, FEM's use $ (N_J, N_e, N, N_d) = (201,64,200,2)$.
 }
\end{table}

\begin{table}
\begin{center}
\begin{tabular}{l l l l}
\hline
 $ l $ & $(101\times8)$ & $(101\times12)$ & $(151\times8)$\\
\hline
  $-5,\dots,5$  & (0.4354, 0.1182) & (0.4362, 0.1176) & (0.4352, 0.1185)\\
  $-5,\dots,10$ & (0.4371, 0.1260) & (0.4378, 0.1253) & (0.4369, 0.1263)\\
\hline
\end{tabular}
\end{center}
\vspace{-2mm}
\caption{Convergence of the PME mode with mesh parameters $(N_J, N_e)$ and cutoff parameters $(l_\textrm{min}, l_\textrm{max})$.
 }
\end{table}

The minimum and maximum radii of the mesh were $ R_\textrm{min} = 0.05 $, and $ R_\textrm{max} = 10$. Doubling or halving $ R_\textrm{min} $ resulted in little change in frequency, whereas larger variations resulted in changes up to 1 per cent or more.

\subsection{The Milky Way model}

The Milky Way Galaxy model includes three components: bulge, halo and stellar disc. Parameters of the components were discussed recently by \citet{GJ13} based on data taken from \citet{S09}.

For the bulge, we adopt a \citet{D93} density profile
\begin{equation}
\rho_b (R) =\frac{\rho_{0}}{y ^{\gamma_b} (1 + y) ^{4 -\gamma_b}}\ ,\quad y =\frac{R}{a_b}\ ,
\label{eq:rho_b_B}
\end{equation}
with spherical radius $ R $, bulge scalelength $ a_b = 0.22 $\,kpc, power-law index $\gamma_b = 0.5 $, scale density $\rho_0 = 336.31M_\odot\textrm{pc}^{-3}$. For the halo density distribution, we assume a cored isothermal profile
\begin{equation}
\rho_h (r) =\frac{\rho_{0}}{1 + y_h ^ 2}\ ,\quad y_h =\frac{\mu}{a_h}\ ,
\label{eq:rho_h_B}
\end{equation}
where $\mu = (r ^ 2 + z ^ 2/q_h ^ 2) ^{1/2}$, the parameter $ q_h = 0.8 $ determines the flattening of the halo, the halo scalelength is $a_h = 3.5$\,kpc, and the central density is obtained from the condition that the density $\rho_h (r_\odot) = 0.01 M_\odot\textrm{pc}^{-3}$ at the solar radius $ r_\odot = 8$\,kpc.

The surface density of the stellar disc has an inner hole, similar to the Besan\c{c}on model \citep{R03}, and is expressed as the difference of two infinitely thin exponential discs
\begin{equation}
 \Sigma_D (r) =\Sigma_{D0} [{\rm e} ^{-r/r_D} -\eps{\rm e} ^{-r/r_{D\eps}}]\ ,
 \label{eq:std_B}
\end{equation}
where $\Sigma_{D0} $ is fixed by the value of the surface density at the solar radius $\Sigma_{D} (r_\odot) = 47\,M_\odot$\,pc$^{-2}$. When $\eps\le 0.4 $ the circular speed of the disc remains growing in the centre, so we took $\eps = 0.4 $. Characteristic scales of the discs are $R_D = 3.0$\,kpc, and $ r_{D\eps} = 1.0$\,kpc. The potential for (\ref{eq:std_B}) can be obtained from the potential of the exponential disc, $\Sigma_{D}\exp (-r/r_D)$ \citep[e.g.][]{BT08}:
\begin{equation}
 V_s (r) = -\pi G\Sigma_D r [I_0 (y) K_1 (y) - I_1 (y) K_0 (y)]\ ,
\label{eq:pot_s_flat}
\end{equation}
where $y = r/(2r_D)$; $ I $, $ K $ are the modified Bessel functions.

\begin{figure}
  \centerline{\includegraphics [width = 85mm]{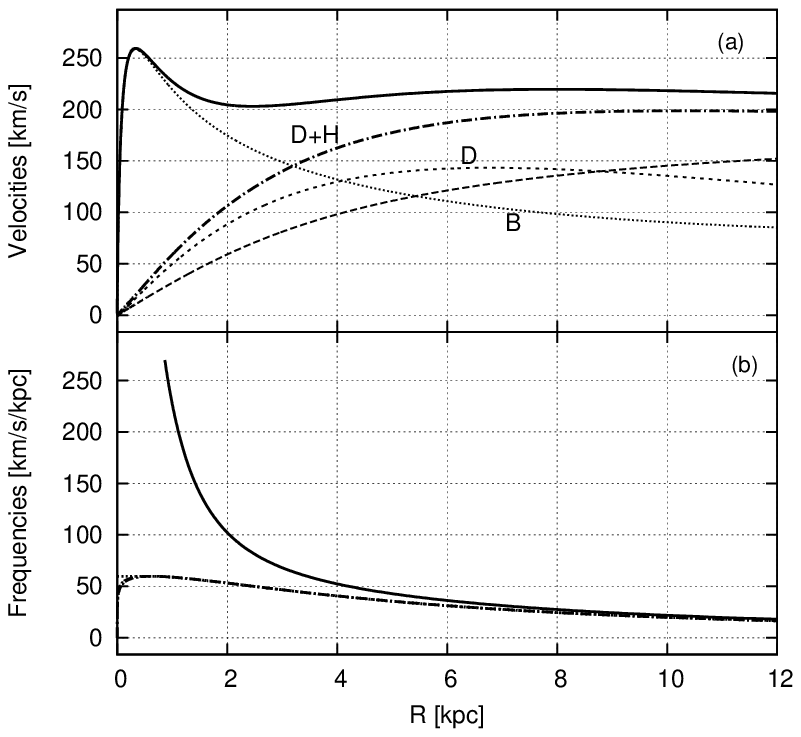}}
  \centerline{\includegraphics [width = 85mm]{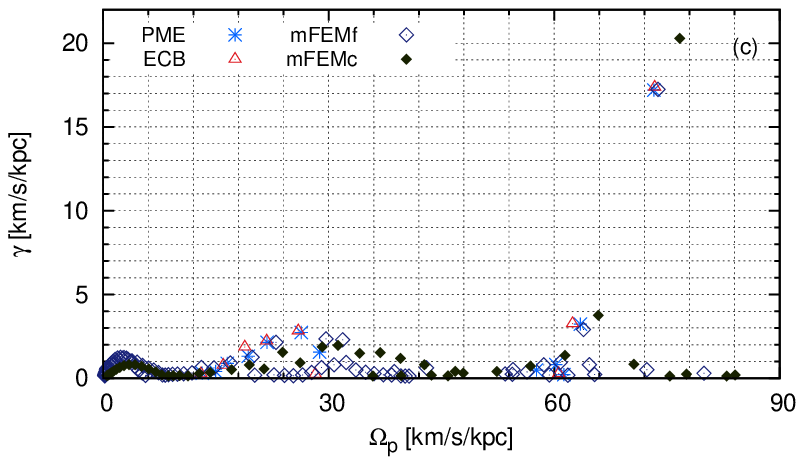}}
  \caption{(a) Circular velocities for the Milky Way model (solid line) and the two-component model without bulge (dash--dotted line). Other curves show the contribution of each component to the rotation curve. (b) Angular velocities $\Omega (r) $, corresponding to profiles (a). Dotted line within $ r = 0.6 $\,kpc shows the corrected angular velocity. (c) Spectra of unstable modes for the two-component model of the Galaxy, obtained by all methods. Modes with growth rates $\gamma < 0.1 $ km/s/kpc are not shown.}
  \label{fig6}
\end{figure}

Fig.\,\ref{fig6} shows the circular velocity profiles for the three-component model and a two-component model without the bulge. The first profile rises steeply in the centre, and gives high angular velocity $\Omega\propto r ^{-\gamma_b/2} $ (see Fig.\,\ref{fig6}(b)). In this respect, the first model is similar to Toomre--Zang model discussed above. In contrast, the two-component model has a much smoother rise in the centre, and hence moderate oscillation frequencies of stars.

Here, we restrict ourselves to the study of the two-component model. Its angular velocity $\Omega (r) $ is increasing within $ r=0.6 $\,kpc, which is unusual for galactic profiles of $\Omega (r)$. Therefore, we have adjusted the frequency and the potential in accordance with a constant value $\Omega (r) = 60$\,km/s/kpc in this region (see dotted line in Fig.\,\ref{fig6}(b)).

A pseudo-isothermal distribution \citep{B11} is used for the unperturbed DF
\begin{equation}
F_0 (J_r, L_z) =\left.\frac{\Omega\Sigma_D}{\pi\sigma_r ^ 2\kappa}\right | _{R_c} {\textrm e} ^{-\kappa (R_c) J_r /\sigma_r ^ 2 (R_c)}\ ,
\label{eq:df_Binney}
\end{equation}
where $ R_c = R_c (L_z) $ is the radius of the circular orbit with angular momentum $ L_z $, and $\sigma_r (r) $ is the radial velocity dispersion, for which we assume
\begin{equation}
\Sigma_r (r) =\sigma_{r0}\Big (0.1 + [(\Sigma_D (r) /\Sigma_D (r_\odot)] ^ q\Big)\ ,
\end{equation}
where $ q = 0.35 $, $\sigma_{r0} = $ 27.3 km/s. As above, we introduce a cutoff function $H_\textrm{cut} =\tanh (L_z/L_0) $, where $ L_0 = 60$\,kpc\,km/s.

\begin{figure*}
  \centerline{\includegraphics [width = 170mm]{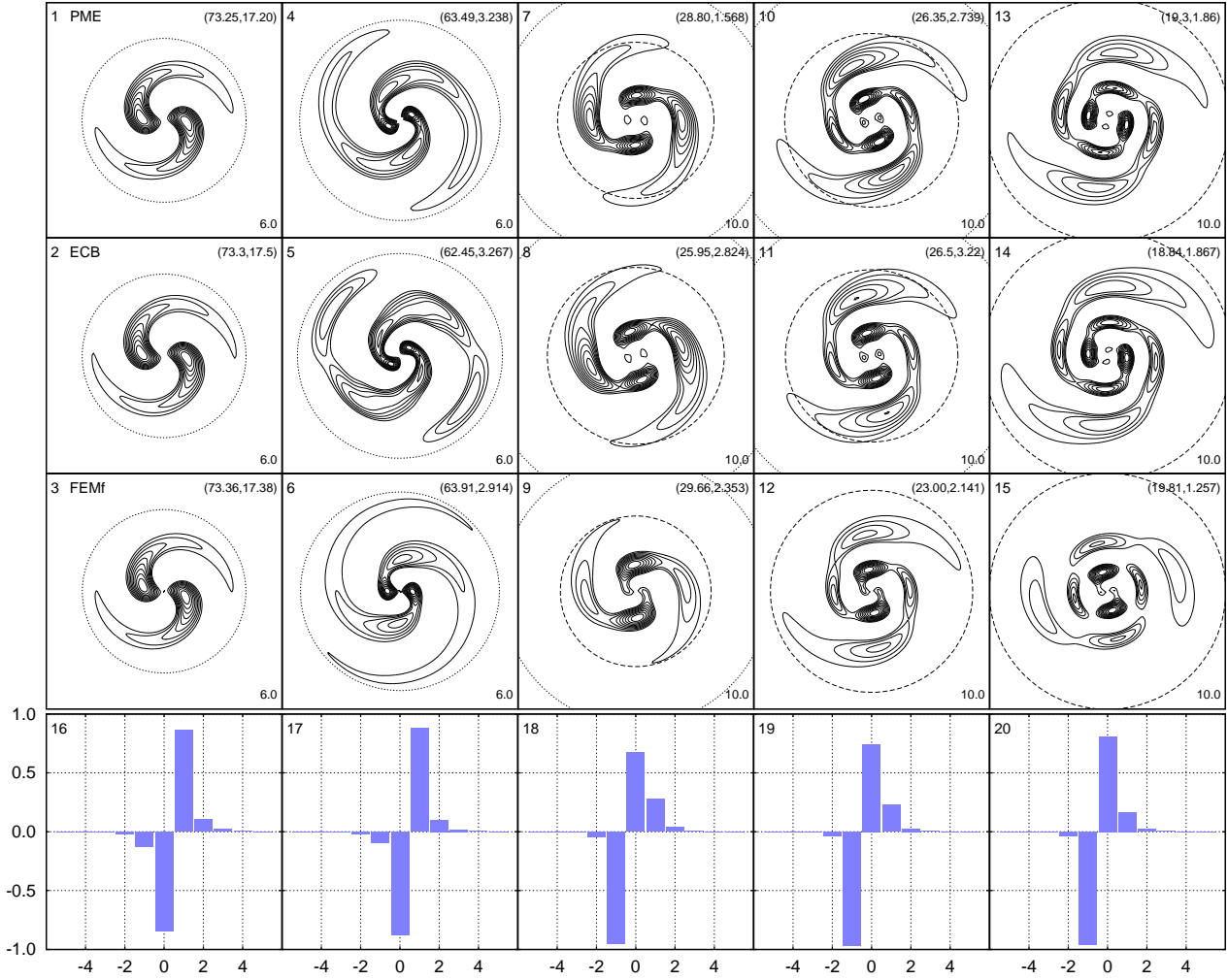}}
  \caption{Same as in Fig.\,\ref{fig3} for the two-component model of the Galaxy, obtained by PME, ECB and FEMf. Frames 1,2,5,6 -- the edge modes; 3,4 -- examples of the patterns for spurious modes of set 1 ($\Omega_\textrm{p} \lesssim 0.4$); 7--10 -- examples of the patterns for spurious modes of set 2 ($\Omega_\textrm{p} \gtrsim 0.65$).}
  \label{fig8}
\end{figure*}

A comparison of the spectra obtained by the different methods is given in Fig.\,\ref{fig6}(c). The PME and ECB spectrum resembles the spectrum of the exponential model, and consist of two well-defined groups of unstable modes. The first group departs from the bifurcation point $\Omega_\textrm{p}\approx 60$\,km/s/kpc, which corresponds to the maximum value of the angular velocity. The second group departs from $\Omega_\textrm{p}\approx 7$\,km/s/kpc, corresponding to the maximum of the curve $\Omega -\kappa/2$. Angular momentum transfers from centre to periphery. Since modes of the first branch lack CR, $l=0$ Fourier component is negative, see Fig.\,\ref{fig8}. Modes of the second branch have positive $l=0$ Fourier component. Note that here we obtained no unstable modes with ILR. The most unstable mode has a frequency $\omega = 146 + 17.3i $ km/s/kpc, corresponding to an e-folding time of the growth rate of $\gamma^{-1} = 57.8$\,Myr.

Contrary to PME and ECB, the FEM methods with interpolating functions (\ref{J10Phi}) completely fail for this model: the outcome consisted of neutral modes only. The reason was in vanishing of the denominator in equation (\ref{J10Phi}) for orbits within $ r = 0.6 $\,kpc. Due to the harmonic potential in this region arbitrary orbits satisfy resonance conditions for purely radial orbits, $ 2l + m = 0 $.
Investigation of $m=1$, for which the resonance structure of orbits is different, confirms this conjecture: FEMf gives $\omega = 196.4 + 20.4i $ km/s/kpc, while PME and ECB give $194.6 + 20.8i $ km/s/kpc, and $195.1 + 20.9i $ km/s/kpc, correspondingly.

To avoid zero denominators, we used an alternative form of interpolating functions
\begin{equation}
    E_l ^ j (n,\vJ) = \frac{F'_{0, l} (\vJ)}{l\Omega_1 + m\Omega_2 + i\gamma_0}\Psi_l ^ j (n,\vJ)\ ,
\label{PhiFEM}
\end{equation}
where $ i\gamma_0 $  is a small imaginary part. The modified FEMf (referred as mFEMf) gives satisfactory results for the most unstable mode, $\omega = 148 + 17.3i $ km/s/kpc; the next fastest growing mode is also close to the corresponding modes of PME and ECB. Modified FEMc (mFEMc) overestimates the growth rate of the most unstable mode by 20 per cent, while for other modes the results are in agreement with mFEMf. Note that the modification distorts the shape of the spectrum near the $x$-axis (see, for example, a series of modes near $\omega = 0 $ in Fig.\,\ref{fig6}(c)).

\section{Conclusions}

This paper presents a comparison of methods for finding unstable modes of stellar discs in the framework of linear perturbation theory. All considered methods are linear eigenvalue problems, allowing us to determine immediately the entire spectrum of unstable modes. A comparison was carried out using three models, which enables us to analyse  advantages of these methods.

Unstable modes of the exponential disc with a soft-centred potential were identified equally well by all methods, except for the simplified finite element method \citep[FEMc;][]{J10}. FEMc overestimates the frequency of the most unstable mode (plus 20 per cent for the growth rate, several percent for the pattern speed). All methods, including FEMc, reproduce well  less unstable modes: the bifurcation points are clearly seen, and the numerical values of the frequencies are close.

The second model (Mestel disc) differs essentially from the others by its flat circular velocity profile in the centre, which gives indefinitely growing, $\propto r ^{-1}$, frequencies of stellar oscillations. The determination of a single unstable mode known with very high accuracy from other work, proved to be a very useful test for the linear matrix methods. In particular, it was found that the ECB method \citep{J07} based on expansion in a Clutton-Brock potential--density basis \citep{CB72} is not reliable here. Other methods, in addition to the fundamental mode, give multiple spurious modes that have noticeable growth rates. The latter can be reduced by using more accurate meshes in action sub-space (PME) or radial rings (FEM).

The collocation scheme \citep{EP05} is fully reliable for the determination of the fundamental mode, despite the restrictions on the mesh parameters in action sub-space. Although the scheme sometimes `adjusts' itself to satisfy mathematical relations with the expense of compromising the local continuity and differentiability of to-be-determined functions, the obtained patterns are even smoother than FEM ones. The best results are within 0.5 per cent for the pattern speed, and 0.8 per cent  for the growth rate from the test values, and a change of the mesh settings indicates convergence to the test values. The finite elements methods \citep{J10} for this model give opposite results compared to the first model: FEMc modes were significantly closer to the test values, while FEMf overestimates the pattern speed and the growth rate by 6 and 13 per cent respectively. Note also that good agreement to the frequency of the test mode was obtained by \citet{J10} for FEMc using an insufficient interval $ -2\le l\le $ 5. We obtained that FEMc values $\Omega_\textrm{p} $ and $\gamma $ converge to the values slightly higher than Toomre's test values (see Table 1).

For the third model, we adopted parameters of disc and halo component from the three-component model of our Galaxy by \citet{GJ13}. The angular velocity profile for the two-component model appeared to be growing in the centre of the disc, so we slightly changed the potential by replacing it with a harmonic one, so that the specified profile became non-increasing. Such a correction should not cause problems, since it affects only a small, hot, and therefore stable region in the centre of the disc. However, this correction led to fail of FEM with interpolating functions proposed by \citet{J10}, which proved to be sensitive to the presence of regions with resonant orbits. Adding an imaginary constant to the denominator of the interpolating functions resolved the problem for the discrete unstable modes with high growth rates.

In the three-component model with a bulge any reasonable value of the pattern speed has the ILR. Since this resonance prevents unstable modes, it needs to be isolated. A possible way is to cutoff the stellar disc between the ILR and CR, or to consider a high $Q$-barrier further out from the ILR \citep{T77, B89}. A detailed investigation of this model will be carried out in a separate paper.

Summarizing, we note that despite the significant progress in the optimization of linear matrix methods made by \citet{J07,J10}, these methods should be used with care. In particular, the applicability of the ECB is limited by the choice of available basis functions. Besides, the presence of resonant orbits may be crucially important for methods that employ interpolating functions in action sub-space (ECB and FEM). Note also that FEM methods suffer from systematic overestimates of frequencies. Referee of the paper pointed out to an `increased rigidity' phenomenon well-known in any finite element analysis. Nevertheless, we anticipate that FEM can be substantially improved by the use of more sophisticated interpolating functions.

\section*{Acknowledgments}

This work was supported by Sonderforschungsbereich SFB 881 `The Milky Way System' (subproject A7) of the German Research Foundation (DFG) and by Programs of Presidium of Russian Academy of Sciences No 17 `Active processes in galactic and extragalactic objects'. We are grateful to M.A. Jalali for discussion and help in verifying our code.

\end{document}